\documentclass[a4paper,12pt]{article}
\usepackage[utf8]{inputenc}
\usepackage{epsfig}
\usepackage{amsmath,amssymb}
\usepackage{textcomp}

\newcommand{\be}{\begin{equation}}
\newcommand{\ee}{\end{equation}}

\newcommand{\LL}{\mathcal{L}}
\newcommand{\spmatrix}[1]{\left( \begin{smallmatrix}#1\end{smallmatrix}\right)}
\title{Standard Model with gauge inert variables}
\author{Srijit Bhattacharjee\footnote{email: srijitb@iitgn.ac.in}\\Discipline of Physics \\Indian Institute of Technology Gandhinagar\\ Ahmedabad, Gujarat- 382424, India.}

\begin{document}

\maketitle

\begin{abstract}
The Standard Model of electroweak interactions has been recast as a gauge free theory where the fields present in the Lagrangian are made inert under $SU(2)_L\times U(1)_Y$ gauge transformations. Furthermore, the residual $U(1)_{em}$ gauge freedom of the dynamical fields are also removed to yield a completely gauge free theory. This formulation thus in turn resolves the issue of gauge dependency of the Standard Model effective potential which plays significant role in determining the instability scale of the electroweak vacuum.
\end{abstract}

\section{Introduction}
The most significant discovery in high energy physics during the last decade of twenty first century is the detection of the Higgs boson of mass $\approx 126$ Gev in the Large Hadron Collider \cite{ATLAS, CMS}. The experimental evidences strongly suggest that it is indeed the Standard Model (SM) Higgs (see \cite{Ellis} and the references therein). However a crucial question would be, whether this value is compatible with the requirement that SM Higgs vacuum remains stable if we higher the energy scale upto the scale where effects of new physics will turn on. The analysis of vacuum stability or metastability to get the lower bound of Higgs mass involves computing Effective Potential a la S. Coleman and E. Weinberg \cite{Coleman-Wein}. Since Effective Potential (EP) is the generating functional of zero momentum one-particle irreducible Green's functions, in gauge theories it is plagued by gauge ambiguities. This raises question on the physicality of the results obtained directly from the EP. The issue of gauge 
dependence of EP and it's consequence in the
context of vacuum stability of SM have been investigated by several authors \cite{Jackiw:74, DJ, DeWitt, VD, Kang, Fischler, Frere, Fukuda, Aitchison, Ramaswamy, Piguet, Nielsen, Lin, LW, BLW, Lin-chy, Baacke, Lim, Luca}. Efforts have been made to render the one-loop EP gauge independent for many limited/shorter versions of SM but to our best knowledge no gauge independent analysis been done so far for full SM. This motivates us to look into methods which will furnish the EP for full SM free of gauge ambiguities and make the conclusions drawn from EP most reliable. The problem of gauge dependence is two fold. First, in the process of quantization, we have to introduce gauge-fixing terms, this makes the resulting effective action gauge dependent. This kind of gauge dependence of the EP was first pointed out by Jackiw \cite{Jackiw:74} for scalar electrodynamics. In a later work by Dolan and Jackiw \cite{DJ}, the EP of scalar
QED was calculated in a set of $R_{\xi}$ gauges. It was concluded that only
the unitary gauge  corresponding to a limiting value of the gauge parameter $\xi$ gives sensible results for radiatively induced masses.

In general the vacuum stability analysis performed
in the literature usually focuses in a particular radius in
field space $\phi_{max}$ where $\phi_{max}$ is the threshold value of triggering instability. Since EP is gauge dependent, it is possible that for one
choice of gauge the potential may satisfy the stability requirement below $\phi_{max}$, but for another choice of gauge the potential may become unstable \cite{Lim}.

The difficulty of gauge dependence of quantum effective potential was more or less resolved by the work of Nielsen \cite{Nielsen}. The observables of a theory with radiatively induced symmetry breaking are found to be gauge invariant, if a change in the gauge parameter is accompanied by a suitable change in the ground-state expectation value of the scalar field. Later, Kobes et al. \cite{KKR} generalized this work to include the gauge dependence of the full effective action at zero and finite temperature. However, a choice of the multi-parameter gauge \cite{DJ} such as: $L_{gf}=-{1\over 2 \xi}(\partial_{\mu}A^{\mu}+\sigma \phi_1 +\rho \phi_2)^2$ would break the homogeneity of Nielsen's identity \cite{Lin-chy} and EP calculated in this gauge would have no physical significance. Thus only for a class of gauges, the on-shell value of EP is gauge fixing independent when calculated in a self consistent approximation scheme.

In this article, we reformulate the Standard Model (SM) in terms of manifestly gauge-inert variables. This reformulation is motivated from earlier attempts to make the Weinberg-Salam model free of gauge degrees of freedom \cite{ild, Mcmullan, masson, chernodub, fadd, sbpm1, sbpm2}. This method is different from the non-covariant approach based upon {\it physical EP} constructed as the expectation value of the Hamiltonian in physical states \cite{BLW, BBHL}. We haven't followed the Nilsen's recent procedure to fix the problem of gauge dependence developed in \cite{Niel}. Here, we have first recast the Weinberg-Salam model in terms of manifestly $SU_{L}(2)$ gauge inert fields. Then the remaining $U_{em}(1)$ gauge freedom is also removed using the physical {\it transverse} vector potential introduced earlier in \cite{sbpm1, sbpm2} i.e. with a massless vector boson which is space-time transverse by definition. Therefore, EP of SM now becomes completely free of any gauge ambiguity since there is essentially no 
need to fix any
gauge or go through the Faddeev-Popov procedure. However, it is known that the EP may depend upon the choice of parametrization of the theory. To make it also invariant under field reparametrization one must invoke the Vilkovisky-DeWitt (VD) geometric method. This will ensure the uniqueness of the results concluded directly from EP in SM. In this article we will show that VD method is implemented implicitly while computing the effective action with the redefined fields.  Another salient feature of this reformulation is the gauge group volumes at the quantum level are factored out from the path integral without any gauge fixing. Thus with this alternative description we should have a better control over the encumbrances entailing gauge fixing in SM.  

The organisation of this article is as follows: In the next section we review and recast the full SM in terms of $SU_L(2)\times U(1)_Y$ gauge inert variables by introducing field redefinitions. Section 3 is devoted to remove the residual $U(1)_{em}$ gauge degrees of freedom from the already reduced theory. Section 4 has a brief account on the computation of EP for the complete gauge free SM and it's consequences on vacuum instability. Finally, we conclude this article discussing several outlooks of this reformulation.

\section{Electroweak Theory in $SU(2)_L$ gauge inert form}
Reformulating gauge theories to avoid hindrances due to gauge degrees of freedom has been started since Dirac's \cite{Dirac} proposal of static
electron using nonlocal pre-factors to absorb the $U(1)$ phase
transformation of the bare field. It should also be mentioned that there have been many efforts in the past towards identifying gauge invariant variables and formulating gauge theories in terms of
those \cite {ild, Mcmullan, masson, chernodub, Nielsen, fadd, sbpm1, sbpm2}. Here we have extended the efforts been made so far towards this direction to express the full SM Lagrangian in terms of manifestly gauge inert variables. This section is devoted to review the gauge inert $SU(2)_L\times U(1)_Y$ model which will be starting point of our next section where the residual $U(1)_{em}$ freedom will be reduced.  

To start with we write down the conventional SM Lagrangian:

\begin{eqnarray}
    \LL&=& (D_{\mu}\Phi)^{\dagger}(D^{\mu}\Phi) -\mu^2\Phi^{\dagger}\Phi -\lambda (\Phi^{\dagger}\Phi)^2 \nonumber \\ &-&\frac{1}{4}~tr~
	{\bf B}^2_{\mu\nu}-\frac{1}{4} Y_{\mu\nu}^{2}\nonumber \\ &+&\bar L~i~\gamma^{\mu}(\partial_{\mu}-\frac{ig}{2}{\bf B}_{\mu}+\frac{i~g^{'}}{2}Y_{\mu})L\nonumber \\
	&+& \bar R~i\gamma^{\mu}(\partial_{\mu}+ig^{'}Y_{\mu})R \nonumber \\
	&+& g_Y(\bar L\Phi R +\bar R \Phi^{\dagger} L), \label{lagew}
\end{eqnarray}
where $\mu$ ($\mu^2 < 0$) and $\lambda$ ($\lambda > 0$) are the usual scalar mass and self-coupling constants. ${\bf B}_{\mu\nu}$, $Y_\mu\nu$ are the field strengths of $SU(2)$ and $U(1)$ gauge fields respectively. For simplicity we have only included only one flavor for fermions but this analysis will be equally valid for more than one flavors. $SU(2)$ gauge field ${\bf B}_{\mu} = B_{\mu}^a t^a, a=1,2,3$, the $U(1)$ gauge field $Y_{\mu}$, the Higgs $SU(2)$
doublet $\Phi$, the left-handed $L$ and right-handed $R$ fermions, transforming under $SU(2) \times U(1)$ gauge transformations as:
\begin{eqnarray}
{\bf B}_{\mu} \rightarrow {\bf B}^{(\Omega)}_{\mu} &=& \Omega^{-1} {\bf B}_{\mu}
+\frac{2i}{g} \Omega^{-1}\partial_{\mu} \Omega \nonumber \\
{\bf B}_\mu \rightarrow {\bf B}_\mu^{(\omega)}&=& {\bf B}_\mu \nonumber \\
Y_{\mu} \rightarrow Y_{\mu}^{(\omega)} &=& Y_{\mu} - \frac{2}{g^{'}}\partial_{\mu} \omega \nonumber  \\
\Phi \rightarrow \Phi^{(\Omega)} = \Omega^{-1}\Phi &,& \Phi \rightarrow
\Phi^{(\omega)} =\exp (-i \omega) \Phi \nonumber
\\ L \rightarrow L^{(\Omega)}=\Omega^{-1}L &,& L \rightarrow L^{(\omega)}=\exp (i \omega) L \nonumber \\
R\rightarrow R^{(\Omega)}=R &,& R\rightarrow R^{(\omega)}=\exp (2i \omega) R \label{gtr}
\end{eqnarray}

 where $\Omega$ and $\omega$ are the $SU(2)$ and $U(1)$ gauge transformations respectively. The covariant derivative of the $SU(2)$ doublet scalar field is given by
\be
    D_{\mu}\Phi = \partial_{\mu} \Phi-\frac{ig^{'}}{2} Y_{\mu} \Phi~-{\bf B}_{\mu} \Phi
	\ee
with
$ t^{a}$'s are the Pauli matrices,
$ g $ and $ g' $ and $g_Y$ are the coupling constants.

 We now consider the following polar decomposition of the Higgs doublet:
 \be
    \Phi(x) = {1\over \sqrt{2}}\rho(x)\begin{pmatrix} \chi_1(x)\\
    \chi_2(x)
    \end{pmatrix}
                                        ,
\ee
where the radial field $\rho$, is a real positive scalar (modulus) field and it is completely gauge-inert. The complex or the `phase' part $\chi=\begin{pmatrix} \chi_1(x)\\
    \chi_2(x)
    \end{pmatrix}$ carries all the gauge transformation properties of $\Phi$. The complex components obey the normalization $\chi_1\bar{\chi}_1~+~\chi_2\bar{\chi}_2=1$.  Furthermore introducing an $SU(2)$ valued matrix $s$,

    \begin{eqnarray}
    s=
\begin{pmatrix}
   \bar\chi_{2} &  {\chi}_{1} \\
    -\bar{\chi}_{1} & {\chi}_{2} ~\label{gee}
\end{pmatrix}
\end{eqnarray}

we can rewrite the Higgs doublet as:

\be
 \Phi={1\over \sqrt{2}}\rho s \begin{pmatrix}
              0\\
              1
             \end{pmatrix}
~\label{Higgsrep}\ee

Under an $SU(2)$ gauge transformation
\be
s \rightarrow s^{(\Omega)} = \Omega^{-1} s . \label{geeg}
\ee
The $U(1)$ transformation of $s$ can easily be found from the basic transformation property of $\Phi$ given in (\ref{gtr}),
\be s \rightarrow s^{(\omega)}=s e^{i\omega t_{3}}\ee
Now we are ready to introduce $SU(2)$ inert gauge bosons via $s$. The new Yang Mills triplet ${\bf W}_{\mu} = W_{\mu}^a t_a$ is defined as
\be
{\bf W}_{\mu} \equiv s^{\dag} \left( {\bf B}_{\mu} + \frac{2i}{g}\partial_{\mu} \right) s
\ee
it is easy to see that under an $SU(2) \times U(1)$ gauge transformations,
\begin{eqnarray}
{\bf W}_{\mu}^{(\Omega)} &=& {\bf W}_{\mu} \nonumber \\
{\bf W}_{\mu}^{(\omega)} &=& e^{-i\omega t_3} {\bf W}_{\mu} e^{i \omega
  t_3} -\frac{2}{g}t_3 \partial_{\mu}\omega . ~\label{ginv}
\end{eqnarray}
Next, we introduce the charged vector bosons by defining
$W^\pm_\mu = \tfrac{1}{\sqrt{2}}(W^1_\mu \mp i W^2_\mu)$, such that

\begin{equation*}
{\bf W}_{\mu} = \begin{pmatrix}
W^3_{\mu} \rule[-10pt]{0mm}{0mm} & \sqrt{2} W^+_{\mu} \\
\sqrt{2} W^-_{\mu} & -W^3_{\mu}
\end{pmatrix},
\end{equation*}
The $U(1)$ transformations of the charged and neutral weak vector bosons can be conveniently represented by the following:
\begin{equation*}
{\bf W}^{(\omega)}_{\mu} = \begin{pmatrix}
W^3_{\mu} -\tfrac{2}{g} \partial_{\mu}\omega t_3  \rule[-10pt]{0mm}{0mm} & e^{-2i\omega t_3} \sqrt{2} W^+_{\mu} \\
e^{2i\omega t_3}\sqrt{2} W^-_{\mu} & -(W^3_{\mu} - \tfrac{2}{g} \partial_{\mu} \omega t_3)
\end{pmatrix},
\end{equation*}

Clearly, the $W^\pm_\mu$ and $W^3_\mu$ fields, are invariant under the action of the gauge group $SU(2)$. However, the $W^\pm_\mu$ fields are $U(1)$-charged with opposite charges, whereas the $W^3_\mu$ fields behave like a $U(1)$-gauge potential.

Now lets check the fermions. Again we use the $SU(2)$ valued matrix $s$ to define $L' = s^{-1}L$ and $R' = R$. The covariant derivatives of left and right handed fermions obviously obey the following transformation rule:

\begin{align}
D^L_\mu L &=s (\partial_\mu - i \tfrac{g}{2} {\bf W}_\mu + i \tfrac{g'}{2} Y_\mu)L'
\nonumber \\
& \equiv s {D'}^{L}_\mu L' \nonumber \\
D^R_\mu R &= (\partial_\mu + i g' Y_\mu) R' = D^R_\mu R' \nonumber \\
g_Y (\overline{L} \Phi R + \overline{R} \Phi^\dagger L) &= {g_Y \over \sqrt2} \rho (\overline{e_L} e_R + \overline{e_R} e_L)\,.
\end{align}

Here, $L'= \spmatrix{\nu_L \\ e_L}$ and $R' = e_R$ or $L'= \spmatrix{t_L \\ b_L}$
and $R' =t_R$ or $b_R $, although the analysis presented here will be valid for any number of generations of leptons or quarks.

It is easy to verify that the individual components of left and right fermions transform as:
\begin{eqnarray}
 \nu_L \rightarrow \nu_L^{(\Omega)}=\nu_L &,& \nu_L \rightarrow \nu_L^{(\omega)}=\nu_L \nonumber \\
 e_L \rightarrow e_L^{(\Omega)}=e_L &,& e_L \rightarrow e_L^{(\omega)}=e^{2i\omega}e_L \nonumber \\
e_R\rightarrow e_R^{(\Omega)}=e_R &,& e_R\rightarrow e_R^{(\omega)}=e^{2i \omega} e_R \label{gtr}
\end{eqnarray}

Now we introduce the neutral massive vector boson and massless photon via usual definitions\\

$\cos \theta_W = \tfrac{g}{\sqrt{g^2 + g'^2}}$ and $\sin \theta_W = \tfrac{g'}{\sqrt{g^2 + g'^2}}$,
\begin{align}
Z_\mu &= \cos \theta_W W^3_\mu - \sin \theta_W Y_\mu
\nonumber \\
A_\mu &= \sin \theta_W W^3_\mu + \cos \theta_W Y_\mu
\end{align}

These gauge bosons have the following transformation rules

\begin{align}
\label{transfZA}
Z_\mu^{(\omega)} &= Z_\mu
&
A_\mu^{(\omega)} &= A_\mu -  \frac{2}{e} \partial_\mu \omega \nonumber \\
Z_\mu^{(\Omega)} &= Z_\mu
&
A_\mu^{(\Omega)} &= A_\mu
,
\end{align}
where the electric charge $e$ is defined as $e=g \sin \theta_W$.

The kinetic part of the neutral and charged vector bosons can be represented as follows:

\be
-\frac{1}{4} Y_{\mu\nu}^{2} - \frac{1}{4} (
	W_{\mu\nu}^{3} + H_{\mu\nu})^{2} - \frac{1}{4}
	(D_{\mu}W_{\nu}^{+} -D_{\nu}W_{\mu}^{+})
	(D_{\mu}W_{\nu}^{-} - D_{\nu}W_{\mu}^{-}),
\ee
    where
\begin{align*}
    W_{\mu\nu}^{3} =& \partial_{\mu} W_{\nu}^{3} - \partial_{\nu} W_{\mu}^{3}\\
    H_{\mu\nu} =& \frac{g}{2i} (W_{\mu}^{+}W_{\nu}^{-} -W_{\mu}^{-}W_{\nu}^{+})
\end{align*}
    and
\be
    D_{\mu} W_{\nu}^{\pm} =\partial_{\mu}W_{\nu}^{\pm} \mp
	igW_{\mu}^{3} W_{\nu}^{\pm} .
\ee

With these redefinitions the complete SM can now be recast as a $SU(2)$ gauge free theory and the resulting Lagrangian is given by

\begin{align}
\label{SMnewv}
\mathcal{L} =&{1\over 2} (\partial_\mu \rho)(\partial^\mu \rho) - {1\over 2}\mu^2 \rho^2 - \frac{\lambda}{4} \rho^4
+ \rho^2(\tfrac{g^2+ g'^2}{8} Z_\mu Z^\mu + \tfrac{g^2}{4}W^+_\mu W^{-\,\mu}) \nonumber \\ +& {1\over \sqrt{2}}\rho g_Y (\overline{e_L} e_R+ \overline{e_R} e_L)
 + \overline{L'} i \gamma^\mu {D'}^{L}_\mu L' + \overline{R'} i \gamma^\mu D^R_\mu R' \nonumber \\
   -&\frac{1}{4}(W_{\mu\nu}^{3})^{2} -\frac{1}{4} H_{\mu\nu}^{2}-\frac{1}{2}W_{\mu\nu}^3H^{\mu\nu} \nonumber \\-&\frac{1}{4}(D_{\mu}W_{\nu}^{+} -D_{\nu}W_{\mu}^{+})
	(D_{\mu}W_{\nu}^{-} - D_{\nu}W_{\mu}^{-})-\frac{1}{4}\frac{g^{'2}}{g^2+g^{'2}} Z_{\mu \nu}^2-\frac{1}{4}\frac{g^2}{g^2+g^{'2}} A_{\mu \nu}^2
\end{align}

 with
\begin{equation*}
    Z_{\mu\nu} = \partial_{\mu}Z_{\nu} -\partial_{\nu}Z_{\mu} , \quad
    A_{\mu\nu} = \partial_{\mu}A_{\nu} -\partial_{\nu}A_{\mu} .
\end{equation*}

In (\ref{SMnewv}) replacement of $W_\mu^3$ with linear combination of $Z_\mu$ and $A_\mu$ has to be made.

\section{Removing $U(1)_{em}$ gauge freedom}
The removal of residual $U(1)$ gauge degrees of freedom from the Lagrangian (\ref{SMnewv}) will follow along similar treatment depicted in \cite{sbpm1, sbpm2}. The essential trick is again to decompose the charged matter fields of the Lagrangian into radial and phase parts and couple those with the {\it physical} part of photon field in a gauge-free fashion. The modulus (radial part) of the matter fields
carries the ``spin'' (scalar) degrees of freedom and the phase part carries only
the ``charge'' of it \cite{chernodub, sbpm1, sbpm2}. This separation of spin and charge actually enables us to represent the theory in terms of manifestly gauge-inert
variables. The photon field being massless classically has only two transverse degrees of freedom. One can decompose the photon field into a transverse and longitudinal part, where the transverse part remains unaltered under a $U(1)$ gauge transformation but the longitudinal or the unphysical part gets changed. The {\it physical} part of the photon is divergenceless by construction and defined as follows: 

\be
 A_\mu=A_{\mu}^T + A^L_\mu=A_{\mu}^T + \partial_{\mu} \int d^4x' G(x-x') \partial' \cdot A(x')
\label{Adecomp}\ee

It is easy to verify from (\ref{Adecomp}) that $\partial.A^T=\partial.(A-A^L)=0$. The longitudinal part of the photon can be expressed as $\partial a(x)$, where the function $a=\int d^4x' G(x-x') \partial' \cdot A(x')$ with $G(x-x')$ being the Green's function for the d' Alembertian. It is easy to see that under standard abelian gauge transformations the $A_\mu^L$ part or the function $a$ changes as:

\be
a \rightarrow a^{(\omega)}=a\,-\,{2\over e} \omega \label{atransf}
\ee
One can capitalize this property of the longitudinal part of the photon field to render the couplings with matter fields gauge inert.

\vspace{0.5cm}

\subsection{Abelian Higgs model}

We now briefly illustrate how this prescription works in case of abelian Higgs model. This description starts with the radial decomposition of complex Higgs field $\phi = (\rho/\sqrt{2}) \exp i\theta$. The modulus part remains unaltered under an abelian transformation but the phase $\theta$ changes as $\theta \rightarrow \theta -2\omega$. The action for abelian Higgs model in this basis now reads (suppressing obvious indices),

\be
S[\rho,\theta, A^T, a] = \int d^4x \left[ \frac12 (\partial
\rho)^2 + \frac12 e^2 \rho^2 (A^T - \partial ( \theta - ea))^2
~- \frac12 (\partial A^T)^2 - V(\rho) \right] 
\ee

Now we can define $\Theta \equiv \theta - e a$ which is manifestly gauge inert,
obvious from the transformations of $a$ and $\theta$. 
Following our prescription above, coupling to the physical photon field is obtained through the action:

\be\label{abh2}
S[\rho,\Theta,A^T] = \int d^4x \left[ \frac12 (\partial
\rho)^2 + \frac12 e^2 \rho^2 ( A^T - \partial \Theta)^2
~- \frac12 (\partial A^T)^2 - V(\rho) \right] 
\ee

Here, $ A^T$ obeys the divergenceless constraint. 
It is interesting that the phase field $\Theta$
occurs in the action only through the combination $A^T - \partial \Theta$; this implies that the shift $\Theta \rightarrow \Theta +
const.$ is still a symmetry of the action. However, since there is no
canonical kinetic energy term for the gauge inert $\Theta$ field it is hard to associate any propagating degrees of freedom with it. In fact we can introduce a new vector boson $Y_\mu=A_\mu^T+\partial_\mu \Theta$ with a {\it physical} longitudinal part $\partial \Theta$ and cast the whole model in terms of $\rho$ and $Y_\mu$ only. Whichever way one may write the important thing is the action has been recast with variables which are manifestly gauge-inert.

\subsection{Electroweak model}
We now turn our attention towards the charged vector fields $W_\mu^{\pm}$. We will perform similar trick as described above. Redefining $W_\mu^{\pm}$ as:

\be
W_\mu^{\pm}=w_\mu e^{\pm \theta^{(\mu)}}~, no~sum~on~\mu
\label{Wdef}
\ee

which implies that under $U(1)$ gauge transformation

\begin{eqnarray}
[w_{\mu}]^{(\omega)} = w_{\mu}~,~[\theta^{(\mu)}]^{(\omega)} = \theta^{(\mu)}
-2\omega ~. \label{wgau}
\end{eqnarray}

As mentioned earlier one can think of $w^{\mu}$ as the component of the charged vector boson carrying only the {\it spin} while $\theta^{(\mu)}$ is the {\it charge} mode. This way of redefining now enables us to cast the kinetic parts of the charged bosons as:
\begin{eqnarray}
D_{\mu} W_{\nu}^{\pm} = \left[\partial_{\mu} w_{\nu} \pm
i w_{\nu}\left( \partial_{\mu}\theta^{(\nu)} +\frac{g^2}{\sqrt{g^2+g^{'^2}}} Z_\mu +e A_{\mu}^T + e \partial_\mu a \right) \right] e^{\pm i\theta^{(\nu)}}
\label{kinW}\end{eqnarray}
Now, from (\ref{wgau}) and (\ref{atransf}) it can be easily verified that the quantity $\Theta^{(\mu)}=\theta^{(\mu)}-ea$ is invariant under $U(1)$ transformations. Therefore,
eq. (\ref{kinW}) is now represented in terms of completely gauge free variables,

\begin{eqnarray}
D_{[\mu} W_{\nu]}^+ D^{[\mu} W^{\nu] -} &=& \frac12\{
[ \partial_{\mu} w_{\nu} \partial^{\mu} w^{\nu} +  w_{\nu} ({\tilde A}^{(\nu)}_{\mu} + {eg \over g'} Z_{\mu}) w^{\nu}( {\tilde A}^{\mu (\nu)}  + {eg \over g'} Z^{\mu})  ] \nonumber \\
&-& \cos \Theta^{(\mu \nu)} [ \partial_{\mu} w_{\nu} \partial^{\nu} w^{\mu} +
w_{\nu} ( {\tilde A}^{(\nu)}_{\mu} +{eg \over g'}Z_{\mu}) w^{\mu} ({\tilde
  A}^{(\mu) \nu} +{eg \over g'} Z^{\nu} ) ]\}~\label{nabs}
\end{eqnarray}

where, ${\tilde A}^{(\mu)}_{\nu} \equiv e A_{\nu}^T + \partial_{\nu}
\Theta^{(\mu)}$ and $\Theta^{(\mu \nu)} \equiv \Theta^{(\mu)} - \Theta^{(\nu)}$.
Since all the components of the $W^{\pm}$ carry same electric charge $\pm1$, one can in fact choose the phases $\Theta^{(\mu)}$ to be the same, independent of the $\mu$. With this choice, eq. (\ref{nabs}) becomes

\begin{eqnarray}
D_{[\mu} W_{\nu]}^+ D^{[\mu} W^{\nu] -} &=& w_{\mu \nu}^2 + \frac 12
w^2 \left( eA^T + \partial \Theta + {eg \over g'} Z \right)^2
\nonumber\\
&-& \frac12 \left [w \cdot \left(e A^T + \partial \Theta + {eg \over g'}
Z \right) \right]^2 ~, \label{nabs0}
\end{eqnarray}

where, $w_{\mu \nu} \equiv 2\partial_{[\mu} w_{\nu]}$.

The other parts involving charged and neutral bosons can be easily made inert under $U(1)_{em}$ transformation. Since

\be
H_{\mu\nu}=\frac{g}{2i} (w_{\mu}w_{\nu} -w_{\mu}w_{\nu})=0
,\ee

we only have the $(W_{\mu\nu}^3)^2$ as non-vanishing from the rest of the terms. Again, $W_{\mu\nu}^3$ can be represented as linear combination of completely $SU(2)_L\times U(1)_Y$ gauge inert $Z_{\mu\nu}$ and $A^P_{\mu\nu}$. Where

\be A^P_{\mu\nu}=\partial_\mu A_\nu^T-\partial_\nu A^T_\mu
\ee

Thus the gauge free pure bosonic part of the SM Lagrangian now becomes:

\begin{align}
\label{SMnewv}
\mathcal{L}_b =& {1\over 2}(\partial_\mu \rho)(\partial^\mu \rho) - {1\over 2}\mu^2 \rho^2 - \frac{\lambda}{4} \rho^4
+ \rho^2(\tfrac{g^2+ g'^2}{8} Z_\mu Z^\mu + \tfrac{g^2}{4}w_\mu w^{\mu}) \nonumber \\    -&\frac{1}{4}(W_{\mu\nu}^{3})^{2} -\frac{1}{4}w_{\mu \nu}^2 + \frac 18
w^2 \left( eA^T + \partial \Theta + {eg \over g'} Z \right)^2
\nonumber\\
+& \frac18 \left [w \cdot \left(e A^T + \partial \Theta + {eg \over g'}
Z \right) \right]^2 \nonumber \\-&\frac{1}{4}\frac{g^{'2}}{g^2+g^{'2}} Z_{\mu \nu}^2-\frac{1}{4}\frac{g^2}{g^2+g^{'2}} (A^{P})^2_{\mu \nu}
\end{align}

Where $W_{\mu\nu}^3$ has to be again replaced by the combination of $Z_{\mu\nu}$ and $A^P_{\mu\nu}$.

For the fermion sector we again redefine the left and right handed fermions in polar basis separating the 'spin' and 'charge' degrees of freedom:

\be
L_a'=e^{i\psi_L^{(a)}}f_L^a~,~R'_a=e^{i\psi_R^{(a)}}f_R^a
\label{ferdef}
\ee

where $a=1, 2$ and for convenience we have introduced component notation for Left-handed and right-handed Weyl spinors. Here again the radial (real) parts of two spinors $f_{L(R)}$ are inert under $U(1)$ transformations but the phase parts change according to:

\be
\psi_L^{(a)} \rightarrow \psi_L^{(a)(\omega)}=\psi_L^{(a)} + 2\omega~, ~ \psi_R^{(a)} \rightarrow \psi_R^{(a)(\omega)}=\psi_R^{(a)}~+~2\omega
\label{fphasegt}\ee

 We now identify $f_L^1\equiv f_L^{\nu}$, $f_L^2\equiv f_L^{e}$, $f_R e^{i\psi_R}=e_R$ for light fermionic doublet or leptons. For quarks similar identifications are obvious. In the following we will illustrate the procedure for lightest leptons only although the formulation is valid for other leptonic flavors and for quarks also. Since neutrinos are charge less, we set $\psi_L^1=0$ and since we don't have any right-handed counterpart of $\nu_L$ we have set $f_R^a=f_R$ and $\psi_R^{(a)} \equiv \psi_R$. However, for quarks we will have both spin-up and spin-down right handed fermions. We now concentrate on the kinetic part of the fermions of the SM Lagrangian ${\cal L}_f$ which now can be rewritten using the definitions (\ref{ferdef}):

\begin{eqnarray}
 {\cal L}_f &=& i \bar{f_L}^e\gamma.\partial f_L^e + i \bar{f_R}\gamma.\partial f_R~-~ \bar{f_L}^e\gamma.\partial \psi_L^e~-~ \bar{f_R}\gamma.\partial \psi_R \nonumber \\&+& i\bar{f_L}^{\nu}\gamma.\partial f_L^{\nu} -e\bar{f_L}^e\gamma.Af_L^e-e \bar{f_R}\gamma.A f_R \nonumber \\&+& {g\over 2} \gamma^\mu\left(\sqrt{2}\bar{f_L}^ew_\mu f_L^{\nu}e^{i(\theta-\psi_L^e)}~+~\sqrt{2}\bar{f_L}^{\nu}w_\mu f_L^ee^{-i(\theta-\psi_L^e)}\right)\nonumber \\&+&{g\over 2}\left(\bar{f_L}^{\nu}C\gamma.Zf_L^{\nu}+C\bar{f_L}^e\gamma.Zf_L^e\right)\nonumber \\ &+&{g'\over2}\gamma^\mu\left(\bar{f_L}^{\nu}SZ_\mu f_L^{\nu}+\bar{f_L}^e SZ_\mu f_L^e+2 \bar{f_R}Z_\mu f_R\right)
\label{newLf}.\end{eqnarray}

Where $S\equiv \sin\theta_w$ and $C \equiv \cos \theta_w$. Also, $\bar{f}_{L(R)}=f_{L(R)}^{\dagger}\gamma^0$. Now, using the same decomposition of the photon field $A_\mu$ we can re-express (\ref{newLf}) as:

\begin{eqnarray}
 {\cal L}_f &=& i \bar{f_L}^e\gamma.\partial f_L^e + i \bar{f_R}\gamma.\partial f_R~-~ \bar{f_L}^e\gamma.\partial \Psi_L~-~ \bar{f_R}\gamma.\partial \Psi_R \nonumber \\&+& i\bar{f_L}^{\nu}\gamma.\partial f_L^{\nu} -e\bar{f_L}^e\gamma.A^Tf_L^e-e \bar{f_R}\gamma.A^T f_R \nonumber \\&+& {g\over 2} \gamma^\mu\left(\sqrt{2}\bar{f_L}^ew_\mu f_L^{\nu}e^{i\psi'}+\sqrt{2}\bar{f_L}^{\nu}w_\mu f_L^e e^{-i\psi'}\right)\nonumber \\&+&{g\over 2}\left(\bar{f_L}^{\nu}C\gamma.Zf_L^{\nu}+C\bar{f_L}^e\gamma.Z f_L^e\right)\nonumber \\ &+&{g'\over2}\gamma^\mu\left(\bar{f_L}^{\nu}SZ_\mu f_L^{\nu}+\bar{f_L}^e SZ_\mu f_L^e+2 \bar{f_R}Z_\mu f_R\right)
\label{gfLf}\end{eqnarray}

where the phases $\Psi_L=\psi_L^{(2)e}~+{e \over 2}a$ and $\Psi_R=\psi_R~+~e a$, are insensitive under $U(1)$ gauge transformation. Also since the left handed fermion has unit charge $|e|$, we can without loss of generality set $\psi_L^e=\psi_L$. It is clear  from (\ref{fphasegt}) that $\theta~-~\psi_L=\psi'$ is inert under $U(1)_{em}$ transformation. The Yukawa interaction part can also be trivially cast in gauge free form:

\be
{\cal L}_Y={1\over \sqrt{2}}\rho g_Y (\overline{e_L} e_R+ \overline{e_R} e_L)={1\over \sqrt{2}}\rho g_Y(\overline{f_L^e}f_R~+~\overline{f_R}f_L^e)
\label{gfY}
\ee

Since the left and right handed electrons are charge conjugate to each other we have set   $\psi_R-\psi_L=0$. Thus the full fermionic part of the SM Lagrangian has also been made free from $U(1)_{em}$ gauge freedom.

Now it remains to specify integration measure for the gauge free Lagrangian for the full Standard Model. The measure in terms of new variables is given by,
\be
d\mu=\rho^3 {\cal D}\rho{\cal D} A^T {\cal D}Z{\cal
D}w{\cal D}\Theta {\cal D}f_L^{\nu}{\cal D}f_L^e {\cal D}f_R{\cal D}\Psi_L{\cal D}\Psi_R {\cal D}\psi'{\cal D}s {\cal D} \omega \delta[\partial.A^T]
\label{measure}\ee
The $SU(2)$ gauge group volume ${\cal D}s$ and $U(1)$ group volume ${\cal D} \omega$ are separated out without gauge fixing. The transverse nature of the vector field $A^T$ has been specified by the delta function. The fields $\Theta$, $\Psi_L$, $\Psi_R$ and $\psi'$ doesn't have any mass term and thus they are the Goldstone modes. The factor $\rho^3$ in the measure is important as it is necessary to maintain the renormalizability of the theory in the gauge-Higgs sector \cite{sbpm2}. 

\section{Effective potential and Vacuum instability}
As mentioned earlier the effective potential is an essential tool to investigate the vacuum structure of any theory. Here we compute the one-loop scalar effective potential from the SM Lagrangian with gauge inert variables. The partition function for SM can now be written :

\begin{align*}
Z=&\int d\mu ~ \exp {i\int d^4x\, ({\cal L}_b + {\cal L}_f+{\cal L}_Y)}
 \end{align*}

From the partition function one usually calculates the effective action $\Gamma$ employing Legendre transformation:
\begin{eqnarray}
\Gamma[\Phi] ~&=&~ -i\ln Z[{\cal J}] ~-~\int d^4 x~{\cal J} \cdot \Phi~ \nonumber \\
\Phi_{\cal J} ~&=&~ {\delta \ln Z[{\cal J}] \over \delta {i\cal J}} ~,
\end{eqnarray}
where, we have collectively labeled all background fields as $\Phi$ and the sources as
${\cal J}$.
With some normalization convention, the effective potential in one-loop approximation may be summarised by the equation
\be
V_{eff}^{1}(\Phi) = V_{tree}(\Phi) + \frac{1}{64\pi^2}STr M^4(\Phi) \ln\left(\frac{M^2(\Phi)}{\sigma^2}\right)
\ee
$M^2(\Phi)$ is the mass matrix in tree approximation associated with the various particles in the theory and $STr$ is the “supertrace,” representing a sum over all the bosonic and fermionic degrees of freedom. It is known that effective action in quantum field theory being an off shell quantity is plagued by gauge ambiguity. Vilkovisky-DeWitt (VD) method is useful for rendering off shell amplitudes not only free from gauge fixing ambiguity but also from the non uniqueness arising due to different parametrization of the fields \cite{VD}. For scalar QED the difference in one-loop effective potential between two different parametrization was demonstrated in \cite{Kun}. This problem has been attacked by VD method to get an unique effective potential for scalar qed in \cite{Kun, sbpm1}. However, it is also known that VD effective action is equivalent to the conventional effective action when Landau-DeWitt gauge is employed (\cite{frad-tseyt}, see the remark made after eq. (18)). In the path integral technique, 
this gauge free SM will have the transversality constraint in it's measure. Thus, without going through the VD technique we can quantize the SM Lagrangian with the constraint inside the measure which
may be regarded as equivalent to fixing of Landau gauge since both are covariant and ghost-free. Therefore the method depicted above has eventually emerged as necessary and sufficient to make the effective action unambiguous and unique.

Now, the unique effective potential for SM can be obtained by conventional functional method with the measure defined by (\ref{measure}) without any gauge fixing. Expanding the Higgs field $\rho$ around a constant background $\rho_0$ we get the contributions from different sectors to the one-loop effective potential

\begin{eqnarray}
V_{eff}^{(1)}(\rho_0)&=& V_{tree}~+~V_{scalar}~+~V_{boson}~+~V_{fermion}\nonumber \\
 &=& {1\over 2} \mu^2\rho_0^2~+~\frac{\lambda}{4}\rho_0^4~+~\frac{1}{64\pi^2}(\mu^2~+~3\lambda\rho_0^2)^2\ln\left({\frac{(\mu^2~+~3\lambda\rho_0^2)}{\Lambda^2}}\right) \nonumber \\
 &+& \left(\frac{3(2g^4~+~(g^2~+~g'^2)^2)}{1024\pi^2}\right)\rho_0^4\ln\left({\frac{\rho_0^2}{\Lambda^2}}\right)~-~\frac{g_Y^{t~4}}{64\pi^2}\rho_0^4\ln\left({\rho_0^2 \over \Lambda^2}\right)
\label{EP}\end{eqnarray}

Here, we have considered the top Yukawa coupling only (with 3 generations of quarks) since the light quarks will have negligible impact on the vacuum stability analysis. In the SM, it is well known that, if the fermion loop contribution
(dominated by the top quark) to the  one-loop potential were large enough then the electroweak vacuum can get destabilized,  due to the existence of a deeper minimum at large $\rho_0$; this phenomenon is translated into a lower limit on the Higgs mass as a
function of the top quark mass \cite{Linde, Sher}. Now we are ready to analyze the Higgs mass bound from Electroweak vacuum stability. Since the effective potential has no gauge parameter dependence, it will be free of the ambiguities related to gauge fixing. It is clear from (\ref{EP}) the negative sign of top Yukawa coupling in the expression of effective potential will be responsible to make the sum of tree level and one-loop effective potential to vanish. The value of the Higgs field at which this happens will be identified as the instability scale.

In the present scenario neglecting the quadratic terms with respect to the quartic interaction terms (for large Higgs field values) we get the approximated value of the field where the instability will set in (at the weak sale):

\be
\rho_0=\sigma\exp{\left(\frac{-8\lambda\pi^2}{B~+~9\lambda^2}\right)}
\ee
Here $\sigma$ represents the weak scale and
\be
B=\left(\frac{3}{16}(3g^4+2g^2g'^2+g'^4)-g_Y^{t~4}\right)
\ee
The above result indicates that for sufficiently large negative $B$, the potential will turn over for a very large value of the Higgs field. Thus vacuum stability analysis should be carried out with the renormalization group improved effective potential \cite{casas} but here we don't make any attempt to carry out the RG analysis. It is also evident that with the gauge free new variables the RG improvement of effective potential will be free of any gauge dependency. However, we leave this task for future communication.

\section{Discussions}
Full Salam-Weinberg electroweak model have been successfully reformulated in terms of manifestly gauge inert variables thus obviating any breakdown of local symmetry to implement the Higgs mechanism. This way of casting the electroweak sector of SM will reduce the algebra of computing scattering amplitudes corresponding various processes in SM and may provide deeper understanding of the true physical degrees of freedoms of SM. The method adopted in this article to calculate effective potential has been compared to Vilkovisky-DeWitt's geometric method already. One can employ VD technique on the $SU(2)_L \times U(1)_{Y}$ inert Lagrangian where there remains only residual $U(1)_{em}$ gauge symmetry to be fixed. This should also provide a gauge free result. Otherwise, one can equivalently employ Faddeev-Popov procedure and invoke Landau-DeWitt gauge to check if it matches with the one with VD technique.

It is important to note that the Gell-Mann-Nishijima relation between electromagnetic  charge, $SU(2)$ diagonal charge operator and the weak hypercharge $Q=t_3\,+~{Y \over 2}$ is  not been shown explicitly in our computation but the treatment adapted here has indeed maintained this relation always. 

 Still many questions have remained to be addressed regarding this analysis. It is known that Ward–Takahashi identities are a consequence of the gauge invariance of the scattering amplitudes. In functional methods it is deduced demanding invariance of the measure under the functional integral under a gauge transformation. Since the path integral measure for this new reformulation is inert under abelian or non-abelian transformations the Ward-Takahasi identities are trivially satisfied. Another interesting study could be the status of anomalies in this new scenario. Chiral anomaly develops because it is impossible to retain both gauge and chiral symmetries at the quantum level (or for renormalization). It is not clear, for a gauge free theory how the anomaly will show up.

We have calculated the one-loop effective potential for the new variable SM which is manifestly gauge independent. We need to use RG improved effective potential to study the electroweak vacuum instability. This will be free from any gauge hindrance now and qualitative analysis must not be too much different from the studies already been made in this field \cite{casas, Miro, Degrassi}. Nevertheless, We would like to report the outcomes of a more accurate analysis of the vacuum stability of SM with this new variable in near future.

\noindent {\bf Acknowledgments}: S. B. thanks Parthasarathi Majumdar, Sudipta Sarkar
and Pritibhajan Byakti for useful exchanges.

\end{document}